\documentclass[twocolumn,amsmath,amssymb,showpacs]{revtex4}

\usepackage[english]{babel}
\usepackage{amsmath}
\usepackage{amssymb}
\usepackage{graphicx}

\newcommand{\ket}[1]{|#1\rangle}

\begin{document}

\title{The Super-Strong Coupling Regime of Cavity Quantum Electrodynamics}

\author{D. Meiser}
\author{P. Meystre}

\affiliation{Department of Physics, The University of Arizona,
1118 East 4th Street, Tucson, AZ 85721 }


\begin{abstract}
We describe a qualitatively new regime of cavity quantum
electrodynamics, the super-strong coupling regime. This regime is
characterized by atom-field coupling strengths of the order of the
free spectral range of the cavity, resulting in a significant
change in the spatial mode functions of the light field. It can be
reached in practice for cold atoms trapped in an optical dipole
potential inside the resonator. We present a nonperturbative
scheme that allows us to calculate the frequencies and linewidths
of the modified field modes, thereby providing a good starting
point for a quantization of the theory.
\end{abstract}

\maketitle

A striking characteristic of cavity quantum electrodynamics (CQED)
is the conceptual simplicity of the systems involved. Typically,
photons in a single cavity mode interact with atoms with a very
small relevant number of internal quantum states
\cite{referencesCQED}. On the experimental side this simplicity
leads to the precise control of most system parameters and to the
laboratory realization of many idealized theoretical models and
Gedankenexperiments. For example, strongly nonclassical states of
the light field such as e.g. number states
\cite{Krause:numberstate,Brattke:numberstates} can be created, the
entanglement between light and atoms can be studied,
 and important questions related to the
quantum measurement process can be addressed. Over the last two
decades experimentalists further expanded the scope of CQED by
achieving increasing control over the translational degrees of
freedom of the atoms via laser cooling and other cooling schemes,
and CQED also plays an important role in quantum information
research.

In the \emph{strong coupling regime} of CQED the coherent
interaction between a single atom and the light field,
characterized by the Rabi frequency $g$, dominates over the
decoherence processes induced by the coupling to the environment,
and characterized by the spontaneous decay rate $\gamma$ and the
cavity damping rate $\kappa$,
\begin{equation}
g>\gamma, \kappa.
\end{equation}

In contrast to these three characteristic frequencies, whose
relative role in CQED has been explored in great detail in the
past, the role of the free spectral range $\omega_{\rm FSR}$ of
the resonator has largely been ignored so far. However, if one
could achieve experimental conditions such that
\begin{equation}
g > \omega_{\rm FSR} \label{ssccriterion}
\end{equation}
the coupled atoms-cavity system would enter a qualitatively new
regime. In this regime the coupling between atoms and light is
strong already during one round trip in the resonator, which
is in contrast to the conventional strong coupling regime where
sufficiently strong coupling is achieved through recycling of 
the light by means of a high Q cavity. Because the spatial
mode pattern inside the resonator is established during one round
trip it is easy to see that in the super strong coupling limit
the atoms can affect the spatial mode structure itself, and not just
the occupation of the modes as is typically the case in conventional
CQED.

The reason why that regime has not been clearly identified
in the past is that $\omega_{\rm FSR}= c/2L$, where $L$ is the
resonator length, is under most circumstances much too large to
lead to significant effects. The single-atom vacuum Rabi frequency
$g$ scales as $1/\sqrt{L}$, and an easy estimate shows that in the
simplest case, $g$ and $\omega_{\rm FSR}$ become comparable for $L
\simeq \lambda^3/\alpha r_0^2$, where $\lambda$ is the wavelength
of the transition under consideration, $\alpha$ is the fine
structure constant, and $r_0$ is a characteristic size of an
electron orbit. Such
resonator lengths are clearly experimentally unrealistic. However,
as we show in this letter, if we relax the condition that $g
\simeq \omega_{\rm FSR}$ at the {\em single-atom} level, this
regime, which we call the ``super-strong coupling'' regime,
is now within experimental reach for modest
numbers of ultracold atoms trapped in the optical lattice formed
by the standing wave inside an optical resonator.

Two important points need to be made at the outset: first, we
emphasize that the super-strong regime can in principle be
achieved independently of whether or not one is in the
(single-atom) strong coupling regime; and second, the situation
that we are considering should not be confused with the more
familiar situations where large optical dispersions, comparable to
or even larger than the free spectral range of the cavity, are
achieved with macroscopic numbers of atoms, as routinely achieved
in laser physics, nonlinear optics and spectroscopy. Rather, the
hallmark of the super-strong coupling regime is that the cavity
resonances are significantly modified from their vacuum form by a
microscopic number of atoms, possibly as few as a few thousands.
In this limit the mode structure depends on the quantized degrees
of freedom of the atoms, and the atoms can become entangled with
the light field in a qualitatively new way: The coupled atoms cavity
system could be in a state $\ket{\psi_{\rm atoms}^{(1)}}\ket{\psi_{\rm light}^{(1)}}+\ket{\psi_{\rm atoms}^{(2)}}\ket{\psi_{\rm light}^{(2)}}$ in which
the two states $\ket{\psi_{\rm light}^{(1)}}$ and $\ket{\psi_{\rm light}^{(2)}}$ correspond to photons with completely different modefunctions.

It is the availability of ultracold atoms confined in optical traps that
makes this new regime of cavity QED possible. We conjecture that
it will find applications in the study of the statistical
properties of quantum-degenerate matter-wave fields, the
generation of entangled optical and matter waves and more generally,
open up new avenues of investigation in CQED.

\begin{figure}
\includegraphics{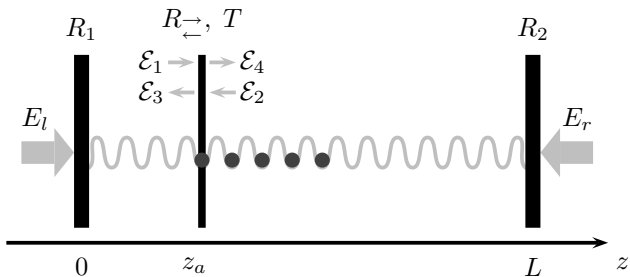}
\caption{Schematic of atoms trapped in an optical lattice
potential in a cavity. Also shown are the effective mirror at
$z_a$ representing the atoms and the boundary conditions for the
incoming and outgoing fields at this mirror.} \label{schematic}
\end{figure}

To set the stage for these future developments, the present paper
is restricted to the classical version of this system. We consider
the specific situation where $N$ two-level atoms with transition
frequency $\omega_a$ are trapped by the optical dipole potential
inside a Fabry-P\'erot resonator with mirror reflectivities $R_1$
and $R_2$ and mirror separation $L$, see Fig. \ref{schematic}. The
$z$-direction is chosen as the optical axis. Two phase-locked
laser beams with frequency $\omega$ and amplitudes $E_l$ and $E_r$
are injected through the left and right mirror, respectively. We
assume that the light is far detuned from the atomic transition
frequency, $|\Delta|=|\omega-\omega_a|\gg \gamma$, so that the
excited atomic state can safely be adiabatically eliminated. In
the far detuned limit the coupling between atoms and field scales
as $1/L$ just like $\omega_{\rm FSR}$ so that the length of the
cavity does not affect their ratio. In the rest of this paper we
measure frequencies in units of $\omega_{\rm FSR}$ so that the
geometry of the cavity becomes irrelevant. Furthermore, if both
the transverse beam profile $u_\perp(r,\varphi;z)$ and the
transverse atomic density profile $\rho_\perp(r,\varphi;z)$ vary
slowly with $z$, the transverse dimensions can be integrated out,
as discussed e.g. in \cite{Meiser:movingmirroradiabatic},
resulting in a one-dimensional effective model.

In this limit the one-dimensional optical dipole potential is
\begin{equation}
V_{\rm dipole}(z)=\Omega(z)|E(z)|^2 \label{dipolepotential}
\end{equation}
where
\begin{equation}
\Omega(z)=\frac{2\wp^2}{\hbar\Delta}A(z).
\end{equation}
In the effective Rabi frequency $\Omega(z)$, $\wp$ is the dipole
moment of the atomic transition, and $A(z)=\int_0^\infty dr
r\int_0^{2\pi} d\varphi |u_\perp(r,\varphi;z)|^2
\rho_\perp(r,\varphi;z)$ is a measure of the overlap between the
atomic density profile and the beam profile and is slowly varying
with $z$. The dipole potential (\ref{dipolepotential}) produces an
optical lattice whose spacing must be determined self-consistently
so that ultracold atoms are trapped at each lattice site. With the
lattice spacing determined in that way, the atoms act effectively
as a microscopic Bragg mirror that scatters the light field
constructively in the \emph{backwards} direction, see Ref.
\cite{Deutsch:optical_lattices}. In this sense the lattice
automatically fulfills a somewhat generalized Bragg condition
corresponding to maximal reflection, regardless of possibly
inhomogeneous occupation numbers at the individual lattice sites.
In the following we therefore assume for simplicity a uniform
filling of $N_s$ sites with $n$ atoms each.

If the local width of the atomic density distribution in each well
is much narrower than an optical wavelength, as will be the case
for a deep optical lattice, we can approximate it as
\begin{equation}
\rho(z)=\sum_{l=0}^{N_s -1} n \delta(z-z_a-l d),
    \label{densitydistribution}
\end{equation}
where $z_a$ is the position of the first occupied lattice site and
\begin{equation}
d=\frac{\pi+2\arctan \Lambda}{k}
\label{dequation}
\end{equation}
is the lattice period. Here $k=\omega/c$ is the wave vector of the
light and
\begin{equation}
\Lambda=(k/4\epsilon_0)n \Omega(z_a)
\end{equation}
is a dimensionless parameter characterizing the collective
interaction between the light and the atoms at a specific lattice
site.

Our starting point for the non-perturbative determination of the
cavity modes is the classical one-dimensional propagation equation
\begin{equation}
\frac{\partial^2 E(z)}{\partial z^2}+k^2 E(z) =
\frac{\Omega}{2}k^2 \rho(z) E(z).
\end{equation}
which can be easily obtained by inserting the polarization in the
far detuned limit, $P(z,t)=-(\Omega/2\epsilon_0) \rho(z)E(z,t)$,
into the Maxwell equations and by invoking the fact that the
atomic density distribution changes very little during one round
trip of the light in the cavity, $\omega_{\rm FSR}^{-1}$.

The problem is greatly simplified by replacing the atomic density
distribution Eq. (\ref{densitydistribution}) by an effective mirror at
a location $z_a$ to be determined later on, with reflection
coefficients $R_\rightarrow$ and $R_\leftarrow$ for the right- and
left-propagating fields and a transmission coefficient $T$, see
Fig. \ref{schematic}.

For the particular density distribution
(\ref{densitydistribution}) the total reflection and transmission
coefficients are readily found by the transfer matrix method as
\cite{Deutsch:optical_lattices,Hemmerich:opticallattices,Slama:Multiplereflections,Meiser:movingmirroradiabatic},
\begin{eqnarray}
R_\rightarrow&=&\frac{-i\Lambda N e^{-2i\arctan \Lambda}}{1-i\Lambda N}\label{reflectioncoefficient}\\
R_\leftarrow&=&\frac{-i\Lambda N e^{-2i\arctan \Lambda(2N-1)}}{1-i\Lambda N}\\
T&=&\frac{e^{-2i\arctan \Lambda N}}{1-i\Lambda N}.
\end{eqnarray}
The steady-state boundary conditions
\begin{eqnarray}
{\cal E}_1&=&T_1 e^{i k z_a} E_l + e^{2ikz_a}R_1 {\cal E}_3,\label{bc1}\\
{\cal E}_2&=&R_2 e^{2ik(L-z_a)}{\cal E}_4+T_2 e^{ik(L-z_a)}E_r,\label{bc2}\\
{\cal E}_3&=&R_\rightarrow {\cal E}_1+T {\cal E}_2,\label{bc3}\\
{\cal E}_4&=&T {\cal E}_1+R_\leftarrow{\cal E}_2,\label{bc4}
\end{eqnarray}
are then easily solved for the field amplitudes at the effective
mirror, see Fig. \ref{schematic}, where ${\cal E}_1$ and ${\cal
E}_3$ correspond to the field amplitudes to the left of the first
atom and the amplitudes at every other atom are easily found using
again the transfer matrix method. From these amplitudes the field
can be determined anywhere inside the cavity through free space
propagation.

In our specific example the mode functions are substantially
altered --- or stated differently the coupling between atoms and
light is of the order of the free spectral range --- provided that
the reflection coefficient $|R_\rightarrow|^2=|R_\leftarrow|^2$ is
of order unity, which is equivalent to $N\Lambda\gtrsim 1$, see
Eq. (\ref{reflectioncoefficient}). There are many possible ways to
achieve such a large value. For example, in the case of
${}^{87}$Rb atoms radially localized much more tightly than the
optical beam waist of $\sim 30\cdot 10^{-6}$ mm, $\wp = 2.32\cdot
10^{-29}$ Cm, $\lambda \simeq 800$ nm, a detuning $\Delta=-10^9
s^{-1}$, we have that $A\approx 3.5\cdot 10^8$ m$^{-2}$ and
$\Lambda/n\approx -9\cdot 10^{-7}$. For a total number of Rb atoms
of $10^6$ we then find $|R|^2\approx 0.45$. All figures in this
letter are for these parameters, together with cavity mirror
reflection coefficients of $|R_1|^2=|R_2|^2=0.99$ and $E_r\equiv
0$. Note that for the case of atoms in an optical lattice the
deviation of $T$ from unity is of the same order in the
interaction as the reflection coefficients. Thus it is an
important feature of the situation at hand that, contrary to the
usual case, it is inconsistent to keep the phase shifts suffered
by the light field upon transmission through the atomic sample
while at the same time neglecting the reflection coefficients.
Finally, we note that appreciable reflection coefficients have
been demonstrated in an experiment by Slama et. al.
\cite{Slama:Multiplereflections} in which reflection coefficients
of atoms in an optical lattice as high as 30\% were demonstrated,
albeit with resonant light.

The mode functions are fully determined by the boundary conditions
Eqs. (\ref{bc1}-\ref{bc4}) once the atomic position $z_a$ is
given. The solutions for the field amplitudes $\mathcal E_i$ are
linear combinations of the injected fields $E_l$ and $E_r$, with
coefficients having resonant denominators given by the determinant
of the set of Eqs. (\ref{bc1}-\ref{bc4}),
\begin{eqnarray}
\label{determinant}
D(\omega)&=&1-R_2 R_\leftarrow e^{2i\omega(L-z_a)/c}- R_1 R_\rightarrow e^{2i\omega z_a/c}\\
&&-R_1R_2(T^2-R_\rightarrow R_\leftarrow)e^{2i\omega L/c}.\nonumber
\end{eqnarray}
The position $\omega_0$ and width $\Gamma$ of the resonances of
the optical cavity dressed by the trapped atoms are given by the
complex zeros of that determinant.

\begin{figure}
\includegraphics[width=8cm]{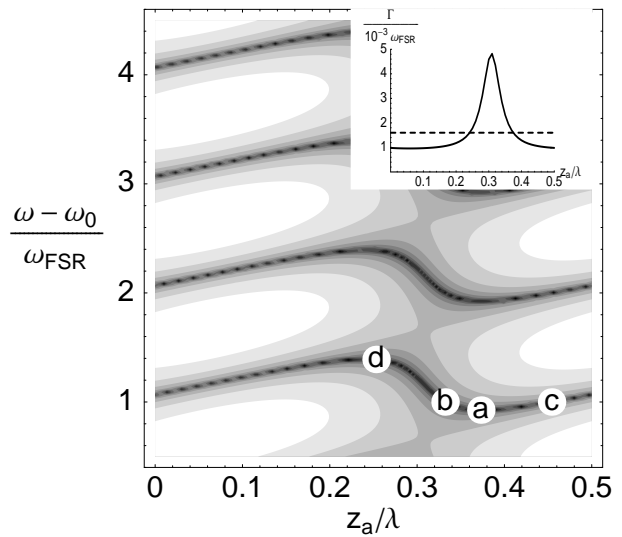}
\caption{Logarithmic plot of $|D(\omega)|^{-2}$ for the parameters
given in the text, with the dark regions corresponding to very small $D$
signifying resonant behavior. The inset shows the linewidth
$\Gamma$ as a function of $z_a$ , with the reflection coefficients
of the atoms taken into account (solid line) and neglected (dashed
line). The labeled dots mark the frequencies and atomic positions for which
the field envelops are shown in Fig. \ref{envelopplot}.} \label{determinantplot}
\end{figure}

Figure \ref{determinantplot}, which shows $1/|D(\omega)|^2$ as a
function of $z_a$ with $\omega$ measured relative to a resonance of
the empty cavity $\omega_0$, illustrates these dressed resonances. The
resonances are associated with strong intracavity fields and
correspond to local minima in the dipole potential for the atoms.
For each atomic position $z_a$ there is an infinite series of such
resonances, separated by the free spectral range $\omega_{\rm
FSR}$. Depending on the atomic positions, the resonances are
shifted by an amount of the order of $\omega_{\rm FSR}$,
confirming that the system is in the super-strong coupling regime.
Furthermore, as illustrated in the inset in Fig.
\ref{determinantplot}, the position of the atoms also affects the
width of the resonances, even in the absence of additional losses
due to spontaneous emission. The resonances can even become
narrower than for the empty cavity, making it very clear that a
simple interpretation of the change in linewidth in terms of
additional loss channels is impossible. The influence of the atoms
on the resonator linewidth is however naturally expected from the
three-mirror cavity model analogy suggested in Fig.
\ref{schematic}, see e.g. \cite{Daendliker:coupledresonators}.

In case the atoms form a uniform gas instead of being located on a
lattice, their reflection coefficient can be neglected and the
determinant Eq. (\ref{determinant}) reduces to $D_{R\equiv
0}(k)=1-R_1 R_2 e^{2i(kL+\phi)}$, where $\phi$ is the phase of
$T$. From this expression it is clear that in that approximation
the interaction with the atoms can only lead to shifts of the
resonance frequencies but not to a change in the linewidths, see
the inset in Fig. \ref{determinantplot}. Also, all changes in the
spectral properties are now independent of the effective atomic
position $z_a$.

\begin{figure}
\includegraphics[width=8cm]{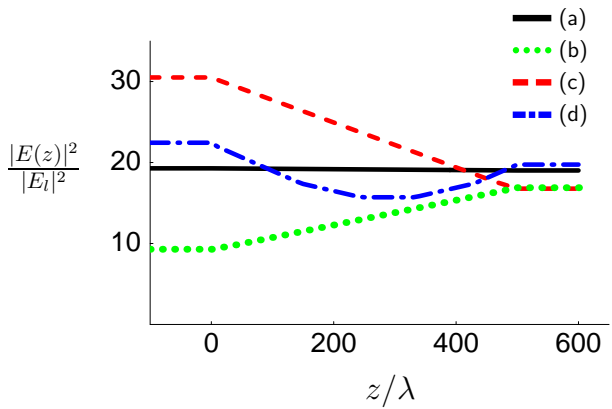}
\caption{(Color online) Field envelopes across the optical lattice
for frequency detunings from an empty cavity resonance of (a)
$\omega = -0.074\omega_{\rm FSR}$, (b) and (c) $\omega = 0$ and
(d) $\omega = 0.39 \omega_{\rm FSR}$ and the respective atomic
positions as shown in Fig. \ref{determinantplot}. For parameters
see text.} \label{envelopplot}
\end{figure}

To confirm that the atoms have a significant effect on the spatial
mode pattern of the cavity, Fig. \ref{envelopplot} shows the field
envelopes along the cavity for three values of the detuning
$\omega$ between the in-coupling light frequency and a resonant
frequency of the empty resonator $\omega_0$, with corresponding atomic
positions $z_a$ labelled by the points (a) through (d) in Fig.
\ref{determinantplot}. (Note that there are two possible values of
$z_a$ for the case $\omega =0)$ In this example, $10^6$ atoms are
distributed over 1000 lattice sites, so that the optical lattice
is approximately 500 optical wavelengths long, and we have set
$E_r\equiv 0$. It is apparent from the figure that the envelope
strongly depends both on the frequency of the incident light and
on the atomic position. Point (a) is at the low frequency edge of
the resonance region, just below the branching point of the
resonance. At this point the field amplitudes at the two edges of
the optical lattice are exactly $\pi$ out of phase (for an odd
number of lattice sites the fields would be exactly in phase with
the same implications) and the field penetrates the optical
lattice almost unperturbed, just as it would in free space
\cite{Deutsch:optical_lattices}. Above the branching point the
field looks entirely different depending on the resonance at which
the atoms are situated. At the left resonance the field is
stronger to the left of the atoms, and at the right resonance it
is stronger on the right. Near the point where the two local
minima merge again (modulo $\lambda/2$) the electrical field
amplitude is extremely sensitive to the frequency of the incident
field as the left and right dominated modes ``collide'' at this
point.

While the quantum dynamics of the coupled atoms-cavity system can
clearly not be understood within the semiclassical description of
the atoms and field considered so far, this approach offers a good
starting point for an effective quantization of the problem: One
can introduce field operators for the self-consistent modes
determined from the boundary conditions (\ref{bc1}-\ref{bc4}),
with frequencies given by the zeros of the determinant
(\ref{determinant}). The modified linewidths shown in the insert
of Fig. \ref{determinantplot} find their physical origin in the
change in the overlap between the cavity field and the continuum
of modes outside the resonator, and can therefore be modelled
using standard quantum optics methods such as e.g. a Born-Markov
master equation.

The atom mirror is in general in a superposition of states with
different reflection coefficients and one could associate
different mode functions with each mirror state. It is worth
noting that the light is only sensitive to the collective
properties of the atoms as represented by these quantized mirrors
and that typically a great number of distinct atomic states give
rise to the same reflection and transmission coefficients. The
number of atomic subspaces that appear indistinguishable to the
light field scales therefore much more favorably with the number
of atoms than the dimension of the total atomic Hilbert space, and
as a result the quantized theory might be simple enough to be
computationally tractable. Based on these ideas we are currently
developing a full quantum theory of the coupled Maxwell and
Schr{\"o}dinger fields. It is expected that it will lead to
fascinating new insight into the dynamics of the coupled
atoms-cavity system that will significantly depend not just on the
internal state, but also the quantum-mechanical center-of-mass
state of the atoms, and will also exhibit significant signatures
of the possible entanglement between the atoms and the light
field.

One difficulty to keep in mind is that in this system the boundary
conditions are dynamical, since the atomic reflection and
transmission coefficients, as well as $z_a$, typically change in
time. As a result, the resonance frequency and the linewidth also
change over time. In general, the quantization of the
electromagnetic field with time-dependent boundary conditions is a
difficult problem, see e.g. \cite{Law:Rubber_cavity_quantization}.
In the present case, we note that, since the spatial mode
structure is established over a time scale of the order of the
round-trip time $1/\omega_{\rm FSR}$, one must have
\begin{equation}
\frac{d |R|^2}{dt}, \frac{d |T|^2}{dt}, \frac{d (z_a/\lambda)}{dt} \ll \omega_{\rm FSR}
\end{equation}
for the above quantization procedure to work. If these conditions
are not satisfied a single-mode theory becomes inadequate and one
must resort to a full multimode description.

In addition to addressing these questions in detail, future work
will investigate modifications in the cooling of the atomic motion
through the inclusion of the atomic reflection coefficient and we plan
to study the effects of the quantized atomic lattice on a moving end mirror
\cite{Vahala:mirroroscillations}.

This work is supported in part by the US Office of Naval Research,
by the National Science Foundation, by the US Army Research
Office, by the Joint Services Optics Program, and by the National
Aeronautics and Space Administration.

\bibliography{mybibliography}

\end{document}